\documentclass[a4paper,twocolumn,11pt]{quantumarticle}
\pdfoutput=1
\usepackage[utf8]{inputenc}
\usepackage[english]{babel}
\usepackage[T1]{fontenc}
\usepackage{amsmath}
\usepackage{hyperref}

\usepackage{tikz}
\usepackage{lipsum}

\begin{document}
	
\title{Adaptive Bayesian algorithm for achieving desired quantum transition}
	
\author{Chengyin Han}

\author{Jiahao Huang}
\address{Guangdong Provincial Key Laboratory of Quantum Metrology and Sensing $ \& $ School of Physics and Astronomy, Sun Yat-Sen University (Zhuhai Campus), Zhuhai 519082, China}

\author{Xunda Jiang}

\author{Ruihuan Fang}

\author{Yuxiang Qiu}
\address{Guangdong Provincial Key Laboratory of Quantum Metrology and Sensing $ \& $ School of Physics and Astronomy, Sun Yat-Sen University (Zhuhai Campus), Zhuhai 519082, China}
\address{State Key Laboratory of Optoelectronic Materials and Technologies, Sun Yat-Sen University (Guangzhou Campus), Guangzhou 510275, China}

\author{Bo Lu}
\email{lubo3@mail.sysu.edu.cn}
\address{Guangdong Provincial Key Laboratory of Quantum Metrology and Sensing $ \& $ School of Physics and Astronomy, Sun Yat-Sen University (Zhuhai Campus), Zhuhai 519082, China}

\author{Chaohong Lee}
\email{lichaoh2@mail.sysu.edu.cn}

\address{Guangdong Provincial Key Laboratory of Quantum Metrology and Sensing $ \& $ School of Physics and Astronomy, Sun Yat-Sen University (Zhuhai Campus), Zhuhai 519082, China}
\address{State Key Laboratory of Optoelectronic Materials and Technologies, Sun Yat-Sen University (Guangzhou Campus), Guangzhou 510275, China}

\maketitle
	
\begin{abstract}
	    Bayesian methods which utilize Bayes' theorem to update the knowledge of desired parameters after each measurement, are used in a wide range of quantum science.
	    For various applications in quantum science, efficiently and accurately determining a quantum transition frequency is essential.
		However, the exact relation between a desired transition frequency and the controllable experimental parameters is usually absent.
		Here, we propose an efficient scheme to search the suitable conditions for a desired quantum transition via an adaptive Bayesian algorithm, and experimentally demonstrate it by using coherent population trapping in an ensemble of laser-cooled $^{87}$Rb atoms.
		The transition frequency is controlled by an external magnetic field, which can be tuned in realtime by applying a d.c. voltage.
		Through an adaptive Bayesian algorithm, the voltage can automatically converge to the desired one from a random initial value only after few iterations.
		In particular, when the relation between the target frequency and the applied voltage is nonlinear, our algorithm shows significant advantages over traditional methods.
		This work provides a simple and efficient way to determine a transition frequency, which can be widely applied in the fields of precision spectroscopy, such as atomic clocks, magnetometers, and nuclear magnetic resonance.
\end{abstract}


\section{Introduction}

The Bayesian approach, which relies on updating the knowledge of the probability distribution after each measurement through Bayes' theorem, can achieve a better precision~\cite{Degen2017,Mehta2019,Fiderer2021,Li2018}.
Bayesian methods have been widely used in quantum information applications and experiments, such as quantum phase/frequency estimation~\cite{Macieszczak2014,Palittapongarnpim2019}, quantum state discrimination or estimation~\cite{Becerra2013,Blume2010}, Hamiltonian learning~\cite{Granade2012,Wang2017} and etc~\cite{Stenberg2014,Granade2016,Ruster2017,Kaubruegger2021,Wang2021}.
%
%
Energy levels are the distinctive property of a quantum system, which can be used as a reference for studying its structure.
Conversely, given the energy levels of a quantum system, one can manipulate the transitions among different levels via applying external fields~\cite{Dowling2003}.
In order to observe and explore most quantum phenomena, certain desired transitions should be achieved, in which the external fields are required to be tuned in a sophisticated manner~\cite{Puebla2020}.

On one hand, the external fields are associated with some specific experimental parameters and their exact relation needs to be determined by fitting a large amount of results.
On the other hand, the windows of the experimental parameters are usually very narrow compared to the adjustable range.
Nevertheless, tuning these parameters in manual makes it inconvenient to find the optimal conditions.
An essential question therefore arises: is there any efficient way to automatically obtain the suitable experimental parameters for a desired transition?

To address this task, machine learning techniques is a natural solution~\cite{Mehta2019,Sarma2019,Fosel2018,Friis2017,Rubio2019,Lumino2018,Valeri2020}.
A possible approach towards this kind of optimization is making use of adaptive protocol~\cite{Lumino2018,Valeri2020,Wiseman1995,Berry2000,Higgins2007,Armen2002,Wheatley2010,Hentschel2010,Daryanoosh2018,Bonato2017}.
In several adaptive protocols~\cite{Fiderer2021,Nusran2014}, Bayesian estimation procedure has been extensively employed.
Bayesian estimation is based on the simple premise that probability distribution can be used for describing uncertainty~\cite{Li2018,Linden2014, Jarzyna2015}.
According to Bayes' theorem, the main features can be inferred effectively through updating the probability distribution after each measurement~\cite{Puebla2020,Wiebe2016,Dinani2019}.
The adaptivity can enhance the measurement precision and save experimental resources compared to non-adaptive ones~\cite{Wang2017,Lumino2018,Valeri2020,Bonato2017,Nusran2014,Dinani2019,Paesani2017}.

As one of the notable phenomena, coherent population trapping (CPT) has been extensively studied~\cite{Zibrov2010,Esnault2013,Liu2017,Breschi2009,Mikhailov2010,Hafiz2017} and widely used for realizing compact high-precision atomic clocks and magnetometers.
The CPT is often produced by a two-photon Raman excitation process.
At the two-photon resonance, atoms are optically pumped into a dark state, which can be used for frequency estimation.
By applying external magnetic fields, the magnetic sublevels are split and one can observe magneto-sensitive CPT signals.
The tunable external fields act the role of additional control parameters, and adaptive protocol provides a tool to search the optimal conditions.

In this article, we propose an efficient scheme to automatically search a desired quantum transition via an adaptive Bayesian algorithm, and experimentally demonstrate it via CPT in an ensemble of laser-cooled $^{87}$Rb atoms.
In our experiment, the desired quantum transition is the magneto-sensitive one between the two hyperfine levels of the $5^2 S_{1/2}$ ground state.
We use the lin$||$lin CPT configuration to realize a high-contrast of dark resonance.
When the frequency difference between the bichromatic field components matches the hyperfine splitting, two-photon dark resonances of $\Lambda$-type systems on the hyperfine levels will be induced.
Here, the two-photon resonance of $|F_g=1, m_F=-1\rangle$ and $|F_g=2, m_F=-1\rangle$ is used and its transition is controlled by an external magnetic field.
Through an adaptive Bayesian algorithm, the voltage (which determines the amplitude of external magnetic field), can automatically converge from a random initial value to the desired one only after few iterations.
We experimentally demonstrate our adaptive Bayesian algorithm and find it is effective and efficient.
If the relation between the target frequency and the controlled voltage is linear, similar to the traditional method, our algorithm can automatically search the desired voltage.
The robustness of our algorithm against noises come from observed CPT spectra as well as the fluctuation of controlled voltage is also demonstrated.
While in the situation where the relation between the target frequency and the controlled voltage is nonlinear, the converged standard deviation obtained via our algorithm show a significant improvement compared to the traditional method.

\begin{figure*}[!htp]
	\includegraphics[scale=0.45]{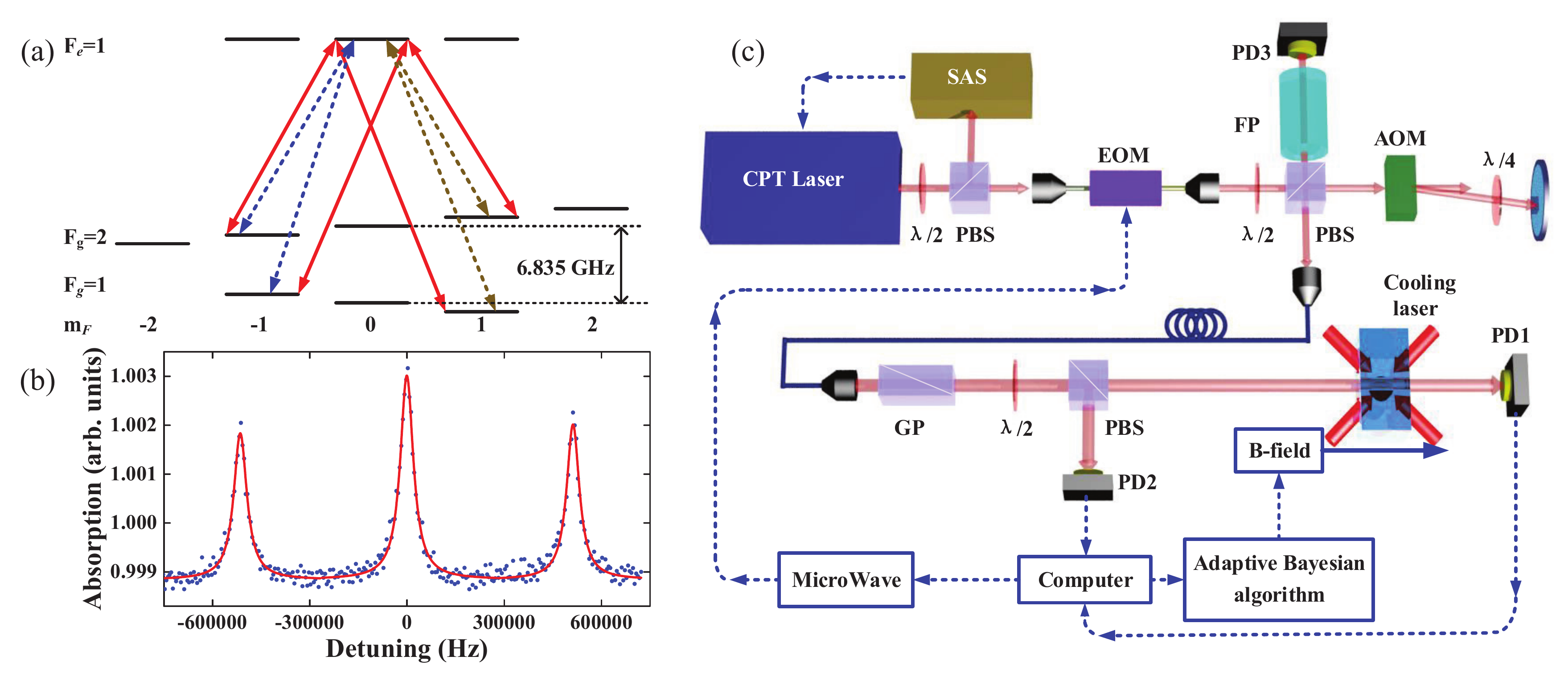}
	\caption{\label{Fig1} Experimental CPT system of laser-cooled $^{87}$Rb atoms.
	(a) Energy levels of $^{87}\textrm{Rb}$ for lin$||$lin CPT scheme. Two separate $\Lambda$ systems of $\sigma^+$ and $\sigma^-$ magneto-insensitive transitions connecting the ground state sublevels $|F_g=1, m_F=\pm1\rangle$ and $|F_g=2, m_F=\mp1\rangle$ with the common excited state $|F_e=1, m_F=0\rangle$ (red solid lines), one $\Lambda$ system of $\sigma^+$ magneto-sensitive transition connecting the ground state sublevels $|F_g=1, m_F=-1\rangle$ and $|F_g=2, m_F=-1\rangle$ (blue dashed lines), and another $\Lambda$ system of $\sigma^-$ magneto-sensitive transition connecting the ground state sublevels $|F_g=1, m_F=1\rangle$ and $|F_g=2, m_F=1\rangle$ (brown dashed lines).
	(b) The observed CPT spectra in the presence of external bias magnetic field.
	(c) Schematic of experimental setup. PBS: polarization beam splitter, EOM: electro-optic phase modulator, PD: photodetector, AOM: acousto-optic modulator, SAS: Saturated absorption spectroscopy, F-P: Fabry-P\`{e}rot cavity, GP: Glan prism. The magnetic field is adapted in realtime through a tunable d.c. voltage controlled by the computer implementing the adaptive Bayesian algorithm.}	
\end{figure*}

\section{ Magneto-sensitive coherent population trapping of Rubidium atoms}

Under the lin$||$lin CPT configuration, the directions of linear polarization of the two CPT frequency components are parallel each other and orthogonal to a tiny static magnetic field~\cite{Zibrov2010,Liu2017,Mikhailov2010}.
Fig.~\ref{Fig1}~(a) shows the coupled energy levels of $^{87}$Rb atoms.
For magneto-insensitive transitions, the bichromatic field simultaneously coupled the atoms via two separate $\Lambda$ systems of $\sigma^+$ and $\sigma^-$ transitions connecting the ground state sublevels $|F_g=1, m_F=\pm1\rangle$ and $|F_g=2, m_F=\mp1\rangle$ with the excited state $|F_e=1, m_F=0\rangle$, respectively.
The first-order Zeeman shifts of the two $\Lambda$ systems have equal strength but with opposite sign.
Both $\Lambda$ systems contribute to the central magneto-insensitive resonance (red solid arrows).
For magneto-sensitive transitions, the bichromatic field accordingly couples the atoms via a $\Lambda$ system connecting the ground state sublevels $|F_g=1, m_F=-1\rangle$ ($|F_g=1, m_F=1\rangle$) and $|F_g=2, m_F=-1\rangle$ ($|F_g=2, m_F=1\rangle$).
Therefore, the CPT resonances yield two microwave magneto-sensitive transitions between $|F_g=1, m_F=\pm1\rangle$ and $|F_g=2, m_F=\pm1\rangle$ (dashed arrows).

We use a five-level model to describe the system and label the five levels as (see Fig.~\ref{Fig1}~(a)):
\begin{eqnarray}
|1\rangle &=& |F=1, m_F=+1\rangle, \nonumber\\
|2\rangle &=& |F=1, m_F=-1\rangle, \nonumber\\
|3\rangle &=& |F=2, m_F=+1\rangle, \nonumber\\
|4\rangle &=& |F=2, m_F=-1\rangle, \nonumber\\
|5\rangle &=& |F'=1, m_F=0\rangle.
\end{eqnarray}
The time-evolution of the system is governed by the Liouville equation for the density matrix~\cite{Shahriar2014},
\begin{equation}\label{Liouville}
\frac{\partial \rho}{\partial t} = -\frac{i}{\hbar}\left(H \rho - \rho H^{\dagger}\right)+\dot{\rho}_{src},
\end{equation}
where the Hamiltonian
\begin{equation}
H=\hbar\left(
\begin{array}{ccccc}
\Delta_1 & 0 & 0 & 0 & \frac{\Omega^b}{2} \\
0 & \Delta_2 & 0 & 0 & \frac{\Omega^a}{2} \\
0 & 0 & \Delta_3 & 0 & \frac{\Omega^b}{2} \\
0 & 0 & 0 & \Delta_4 & \frac{\Omega^a}{2} \\
\frac{\Omega^b}{2} & \frac{\Omega^a}{2} & \frac{\Omega^b}{2} & \frac{\Omega^a}{2} & \delta-\frac{i\Gamma}{2} \\
\end{array}
\right),
\end{equation}
with the parameters
\begin{eqnarray}
\Delta_1 &=& \frac{\Delta}{2} + g_1 \mu_B B_z \nonumber,\\
\Delta_2 &=& \frac{\Delta}{2} - g_1 \mu_B B_z\nonumber,\\
\Delta_3 &=& -\frac{\Delta}{2} - g_2 \mu_B B_z\nonumber,\\
\Delta_4 &=& -\frac{\Delta}{2} + g_2 \mu_B B_z.
\end{eqnarray}
Here, $\Omega^{a}$ and $\Omega^{b}$ are the Rabi frequencies, $\Delta=\delta_1-\delta_2$ and $\delta=(\delta_1+\delta_2)/2$ where $\delta_1$ and $\delta_2$ correspond to the single-photon detuning of the two laser fields. $g_1=-0.5017$ and $g_2=0.4997$ are the effective Land\'{e} g-factors, $\mu_B$ is the Bohr magneton, and $B_z$ is the magnetic field along the $z$ axis that causes the Zeeman shifts. Under the weak magnetic field $B_z$ involved in experiment, the Zeeman sublevels are assumed to undergo only the linear Zeeman shifts.

The term $\dot{\rho}_{src}$ accounts for the influx of atoms into the ground states due to the decay from excited
state, which is defined as
\begin{equation}
\dot{\rho}_{src}= \left(
\begin{array}{ccccc}
\frac{\Gamma_1}{2}\rho_{55} & 0 & 0 & 0 & 0 \\
0 & \frac{\Gamma_1}{2}\rho_{55} & 0 & 0 & 0 \\
0 & 0 & \frac{\Gamma_2}{2}\rho_{55} & 0 & 0 \\
0 & 0 & 0 & \frac{\Gamma_2}{2}\rho_{55} & 0 \\
0 & 0 & 0 & 0 & 0 \\
\end{array}
\right),
\end{equation}
where $\Gamma_1$ and $\Gamma_2$ are the damping rates decaying from the excited states to the ground states $|F=1\rangle$ and $|F=2\rangle$. Generally, $\Gamma_1+\Gamma_2=\Gamma$.
The decay between the ground states are neglected here.

We can solve the time-dependent element $\rho_{ij}(t)$ ($i,j=1,2,3,4,5$) according to Eq.~\eqref{Liouville}.
For $^{87}$Rb atom, the damping rates $\Gamma=2\pi\times6$ MHz, $\Gamma_1=\Gamma/4$, $\Gamma_2=3\Gamma/4$, and the Rabi frequencies $\Omega_a=\sqrt{3}\Omega_0$, $\Omega_b=\Omega_0$ with $\Omega_0=2$ MHz.
Initially, the atoms are equally populated in the four ground states $|1\rangle$, $|2\rangle$, $|3\rangle$, and $|4\rangle$.
The duration of the CPT pulse is set as $\tau=1$ ms, which is sufficiently long to reach the steady state.
$1-\rho_{55}$ characterizes the absorption of light after passing through the atoms.
Through calculating the final population in the excited state $\rho_{55}(\tau)$ for different detuning $\Delta$, we can use the normalized amplitude of ($1-\rho_{55}$) to describe the CPT resonances.
The observed CPT spectra is shown in Fig.~\ref{Fig1}~(b).

Fig.~\ref{Fig1}~(c) depicts the schematic of the experimental apparatus.
The system consists of a three-dimensional magneto-optical trap (MOT), a CPT laser system and an adaptive controller.
The MOT apparatus comprises an ultra-high vacuum cell with pressure of $10^{-8}$ Pa, a quadruple magnetic field produced from a pair of magnetic coils, and laser beams.
Typically, the MOT can trap about $10^7$ $^{87}\textrm{Rb}$ atoms with a 100 ms cooling period.
The CPT beam is generated by a single laser passing through an electro-optic phase modulator (EOM).
The CPT laser source is an ECDL tuned to the $^{87}\textrm{Rb}$ D1 transitions at 795 nm.
The laser beam is split into two parts by a half-wave plate and a polarization beam splitter (PBS).
One beam is used to offset lock the laser frequency to the transition between $|F_g=2\rangle$ and $|F_e=1\rangle$ with saturated absorption spectroscopy (SAS).
The other beam is sent to a fiber-coupled EOM modulated by a microwave at 6.835 GHz.
The  positive first-order sideband forms the $ \Lambda $ systems with the carrier.
The output of the EOM is split into two parts by a half-wave plate and a PBS.
The reflected beam is sent to a Fabry-P\`{e}rot (FP) cavity that monitors the intensity of sidebands generated by the EOM, and the powers of the first-order sidebands are set equal to that of the carrier signal.
Due to far detuned from any resonances, the extra sidebands do not contribute to the CPT signal.
The transmitted beam is sent to a double-pass acousto-optic modulator (AOM) that shifts the optical frequencies to resonances and switches on or off the CPT beam.
Then the beam is coupled into a PMF.
After the fiber, the CPT beam is collimated to an 8-mm-diameter beam and sent through a Glan prism to purify the linear polarization.

\begin{figure}[!htp]
	\includegraphics[width=1.0\columnwidth]{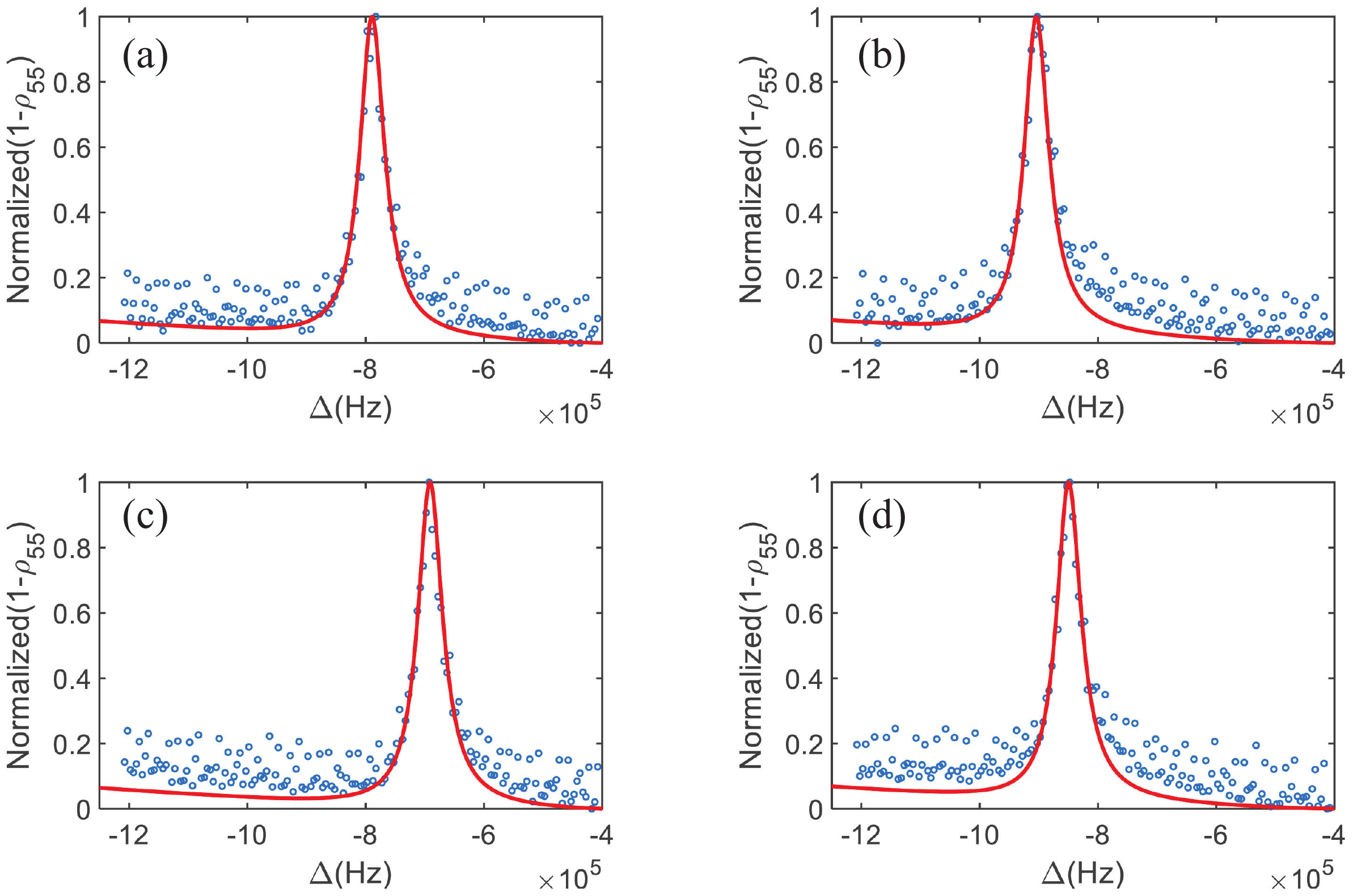}
	\caption{\label{Fig2} Experimental results and numerical simulation of the magneto-sensitive CPT resonance. The observed magneto-sensitive CPT spectra (blue open circles) for (a) $U=8.5000$ V, (b) $U=9.7533$ V, (c) $U=7.4523$ V, and (d) $U=9.1804$ V. These spectra are used as the likelihood function for the iteration with adaptive Bayesian algorithm. The solid lines are the numerical simulation obtained from the five-level model.
	}	
\end{figure}

After the Glan prism, the available laser power is hundreds of microWatt.
Then, the CPT beam is separated equally into two beams by a half-wave plate and a PBS.
One beam is sent to the normalization photodetector (PD2) which is used as a normalization signal to reduce the effect of intensity noise on the CPT signals.
Another beam is sent to interrogate the cold atoms, and the transmitted light is collected on the other photodetector (PD1).
The coherent beam is turned on with 1 ms latency time after turning off the MOT.
The transmitted and normalized beams are synchronously measured with corresponding photodetectors (denoted by $S_T$ and $S_N$) during the CPT pulse.
Finally, the CPT spectra are obtained from the signal $S_T/S_N$.
In order to eliminate the stray magnetic field, three pairs of Helmholtz coils are used for compensation.
To create the bias magnetic field, we employ an additional pair of Helmholtz coils aligned with the direction of the CPT laser beams.
By controlling the coils (whose currents are determined by a d.c voltage $U$), the strength of the bias magnetic field can be precisely tuned.

We focus on the magneto-sensitive CPT resonances.
Since the spectrum is symmetric with respect to $\Delta=0$, we only consider the magneto-sensitive transition in the negative detuning regime.
In Fig.~\ref{Fig2}, the observed experimental results (blue dots) and the corresponding numerical calculations (red lines) for the magneto-sensitive CPT resonance are shown.
The numerical simulation matches well with the experimental results.
In the middle of the peak, the spectrum is in Lorentz lineshape, which can be fitted by a Lorentz function.
The location of the center is the target frequency we desire.
We denote the desired frequency (corresponding to a certain detuning value) as $f_d$, which we need to achieve in experiment.

Generally in experiments, the magnetic field amplitude $B_z$ is proportional to the applied voltage $U$, i.e., $B_z \propto U$.
In order to find out the desired voltage $U_d$ corresponding to the target frequency $f_d$, we can calculate the normalized transmission spectrum for different $B_z$ and extract the relation $f_d=g(U_d)$ numerically.
However, in practice, one should adjust the voltage and measure the corresponding frequency to obtain a relation $f_d=g(U_d)$, and finally determine the applied voltage by the inverse function, i.e., $U_d=g^{-1}(f_d)$.
In order to collect sufficient data for fitting the relation $f=g(U)$, a large amount of center frequency $f$ should be experimentally measured at different values of voltage $U$ in manual, and the value of $f_d$ should be averaged over a large amount of data to reduce the deviation.
Besides, if the relation between the target frequency and the voltage is not linear, this traditional method becomes inconvenient in finding suitable fitting function $g(U)$, as shown in the following Sec. 3.

\section{ Adaptive Bayesian algorithm for searching a desired transition frequency}

In this section, we will show how to automatically obtain the suitable experimental parameters for a desired transition via the adaptive Bayesian algorithm.
We accomplish this task in virtue of an adaptive controller.
The adaptive controller is realized with the help of a computer that includes an adaptive Bayesian algorithm and controls over the experimental parameters via digital I/O devices (NI 6536 and NI 6733).
The bias magnetic field strength is set via a voltage controlled current source and the applied voltage $U$ is controlled by the computer.
The control system is implemented in both Python and LabVIEW which transfer data via TCP sockets.
The LabVIEW program controls the digital I/O devices to obtain the CPT spectra and change the applied voltage value automatically.
The Python program operates the adaptive Bayesian algorithm that updates the applied voltage in realtime.
The adaptive Bayesian algorithm could automatically find the suitable voltage $U_d$ for generating the bias magnetic field corresponding to the magneto-sensitive CPT transition centered at $f_d$.
In the following, we introduce the algorithm procedure in detail and present the experimental demonstration.

\subsection{ Adaptive Bayesian algorithm}

\begin{figure*}[!htp]	
	\includegraphics[width=2\columnwidth]{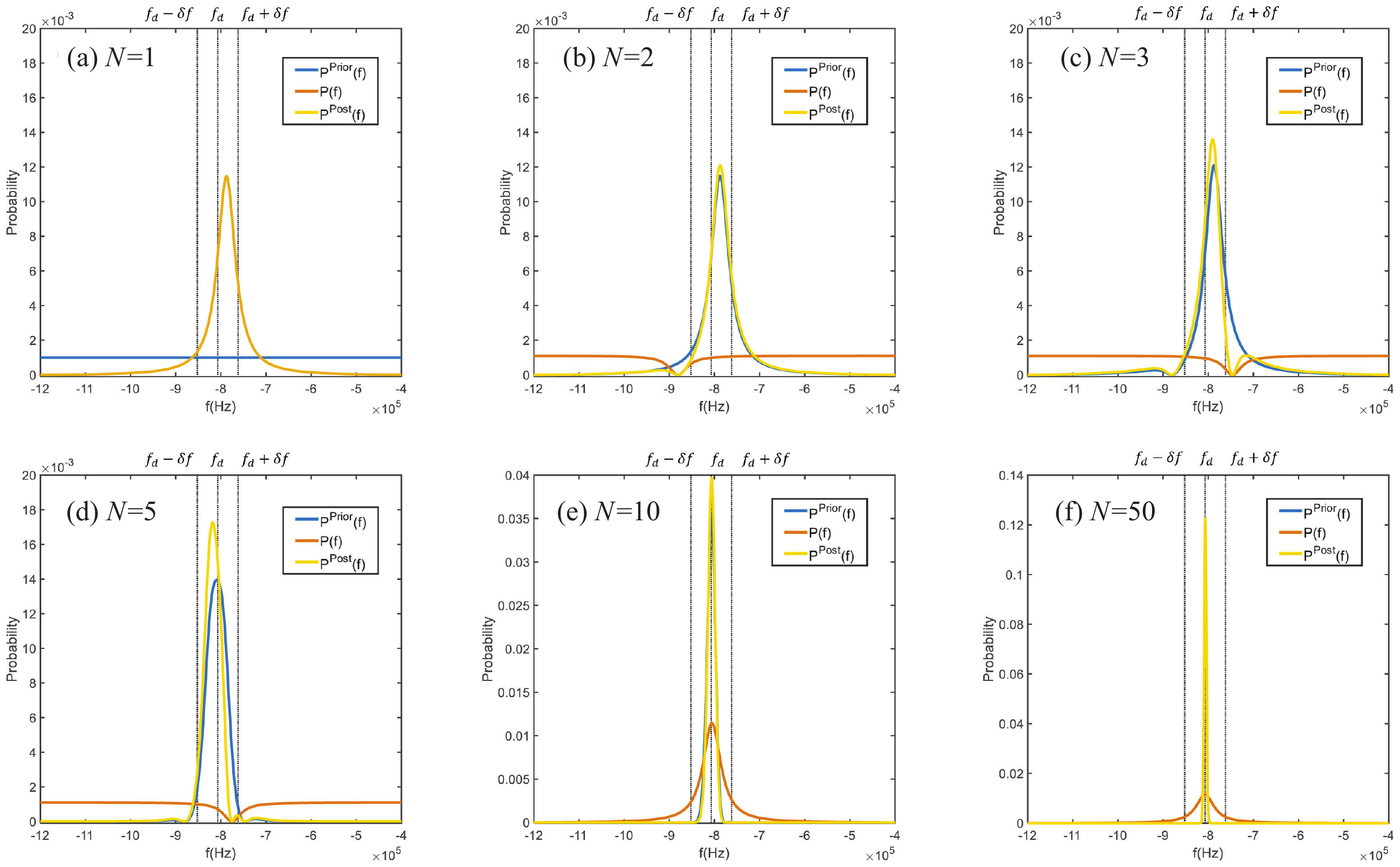}
	\caption{\label{Fig3} Numerical simulation of probability distributions during the adaptation. The blue, orange and yellow lines denote the prior $P^{Prior}_k(f)$, the normalized likelihood $P_k(f)$ and the posterior $P^{Post}_k(f)$ for the $k$-th iteration, respectively. The dashed lines represent the locations of $f_d-\delta f$, $f_d$, and $f_d+\delta f$. Here, the initial voltage $U_0=8.5$ V, the desired magneto-sensitive transition frequency $f_d=-807.6$ kHz. }	
\end{figure*}

Our adaptive Bayesian algorithm for each iteration includes the following key steps~\cite{Lumino2018,Nusran2014}.
\begin{itemize}
	\item \textbf{Step 1}: Input a voltage (the $k$-th voltage is denoted by $U_k$, the initial input voltage $U_0$ is randomly chosen) and sweep the detuning to obtain the magneto-sensitive CPT signal,
	\begin{equation}\label{Lorentz}
	L_k(f)=\frac{2A_k}{\pi}\frac{w_k}{4[(f-\epsilon_k)^2+w_k^2]} + B_k,
	\end{equation}
	which is fitted by a Lorentz lineshape function involving four fitting parameters $A_k$, $w_k$, $\epsilon_k$ and $B_k$.
	
	\item \textbf{Step 2}: Normalize $L_k(f)$ as $\tilde{L}_k(f)$, where $\tilde{L}_k(f)$ becomes ranging from 0 to 1, i.e.,
	\begin{equation}
	\tilde{L}_k(f)=\frac{L_k(f)-\min{L_k(f)}}{\max{L_k(f)}-\min{L_k(f)}}.
	\end{equation}
	Regard $\tilde{L}_k(f)$ as a probability distribution. If the probability at the target frequency $\tilde{L}_k(f_d)>0.5$, the likelihood function remains the same, i.e., $\tilde{L}_k'(f)=\tilde{L}_k(f)$. If the probability at the target frequency $\tilde{L}_k(f_d)\le0.5$, the likelihood function becomes $\tilde{L}_k'(f)=1-\tilde{L}_k(f)$.
	
	\item \textbf{Step 3}: Calculate the normalized likelihood probability distribution,
	\begin{equation}\label{Pk}
	P_k(f)=\frac{\tilde{L}_k'(f)}{\int \tilde{L}_k'(f)\textrm{d}f}.
	\end{equation}
	According to the measured probability distribution $P_k(f)$, obtain the posterior probability distribution,
	\begin{equation}\label{Post}
	P_{k}^{Post}(f)=\frac{P_k(f)P_k^{Prior}(f)}{\int P_k(f)P_k^{Prior}(f)\textrm{d}f},
	\end{equation}
	where the $k$-th prior probability distribution is the $(k-1)$-th posterior probability distribution, i.e., $P_{k}^{Prior}(f)=P_{k-1}^{Post}(f)$.
	Usually, there is no pre-knowledge about the unknown parameter available, the initial uninformative prior is a uniform probability distribution, i.e., $P_1^{Prior}(f)=1/C$ with $C=\int \textrm{d}f$.
	
	\item \textbf{Step 4}: Calculate the mean frequency
	\begin{equation}
	\bar{f}_k=\int f P^{Post}_k(f) \textrm{d}f,
	\end{equation}
	and its standard deviation
	\begin{equation}
	\Delta {f}_k=\sqrt{\int f^2 P^{Post}_k(f) \textrm{d}f - (\bar{f}_k)^2}.
	\end{equation}
	
	\item \textbf{Step 5}: Adjust the voltage of the next step according to the measured probability distribution $P_{k}$.
	
\end{itemize}

The rule for adjusting the voltage at each step is given as
\begin{equation}\label{Uk}
U_{k+1} = U_{k} + (-1)^{s}\cdot h \cdot \delta U/k,
\end{equation}
where
\begin{eqnarray}
s=\left\{
\begin{array}{lr}
0, & \text{if } P_{k}(f_d-\delta f)-P_{k}(f_d+\delta f)<0\\
1, & \text{if } P_{k}(f_d-\delta f)-P_{k}(f_d+\delta f)>0
\end{array}
\right.
\end{eqnarray}

\begin{equation}
h=\frac{P_{k}(f_d-\delta f)-P_{k}(f_d+\delta f)}{P_{k}(f_d-\delta f)+P_{k}(f_d+\delta f)-2P_{k}(f_d)},
\end{equation}
and $\delta U$ is a variation step for the voltage, $\delta f$ is a frequency deviation from the desired frequency $f_d$.

Here, the voltage for the next iteration is adapted according to the current likelihood.
The direction of the variation is determined by $s$ which is related to the two probabilities located oppositely respect to $P_k(f_d)$, and the frequency deviation $\delta f$ can be feasibly chosen. $h$ reflects the competition between the difference $P_{k}(f_d-\delta f)-P_{k}(f_d+\delta f)$ and the gradient $P_{k}(f_d-\delta f)+P_{k}(f_d+\delta f)-2P_{k}(f_d)$.
If the mean of the current likelihood is far from $f_d$, $h$ becomes large and vice versa.
In practice, $h$ is restricted in a modest range, i.e., $h_{min}\le h\le h_{max}$. The variation step $\delta U/k$ is decreased as $k$, and $\delta U$ can be chosen feasibly in experiment.
This rule guarantees the efficiency of our adaptation.

\begin{figure*}[!htp]
	\includegraphics[scale=0.5]{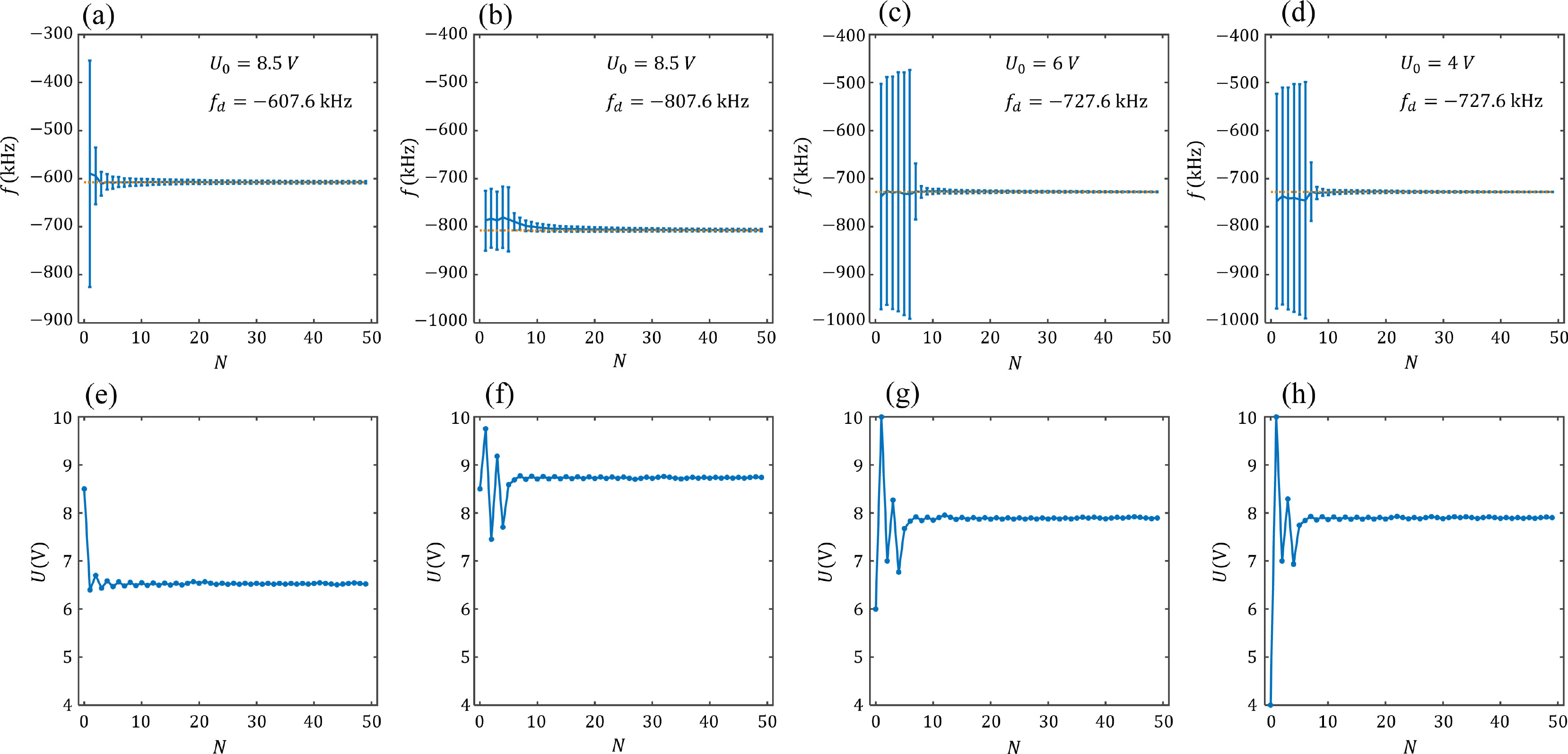}
	\caption{\label{Fig4} Experimental results with adaptive Bayesian algorithm.
		Above row (a)-(d): The converged frequency versus the iteration number for different target frequency $f_d$ with different initial voltage $U_0$. Here, the orange dashed lines denote the target frequency $f_d$. The errorbars correspond to the standard deviations of the converged frequency.
		Bottom row (e)-(h): The corresponding variation of the applied voltage versus the iteration number $N$ for the above row. The voltage for the next step is adapted according to the Bayesian algorithm. If the estimated frequency is far away from the target, the variation amplitude of the voltage for the next step will get large, and vice versa. Comparing (a, e) and (b, f) for the same initial voltage, the final voltage will converge to the corresponding value according to the target frequency.  Comparing (c, g) and (d, h), despite the adaption of voltage is different, the final stages of the iteration are similar and the suggested voltage is the same for $U_0=6$ V and $U_0=4$ V.}	
\end{figure*}

When $k$ is modestly large, the mean frequency $\bar{f}_k$ will converge to the desired frequency $f_d$ with reduced standard deviation $\Delta f_k$. Meanwhile, the converged voltage will be close to the desired condition for achieving the magneto-sensitive transition with frequency $f_d$.
Finally, the average of the converged voltage for the last few iteration number can be given as the suggested applied voltage $U_d$.

In Fig.~\ref{Fig3}, we perform the numerical simulation to see how it works.
First, the initial voltage is chosen as $U_0=8.5$ V.
The prior for the first iteration $P_1^{Prior}(f)$ is a flat distribution, see the blue line in Fig.~\ref{Fig3}~(a).
We obtain the magneto-sensitive CPT signal and normalize it as the likelihood $\tilde{L}_k(f)$.
Since $\tilde{L}_k(f_d)>0.5$, the likelihood remains as $\tilde{L}_k'(f)=\tilde{L}_k(f)$.
According to Eq.~\eqref{Pk}, we calculate the normalized likelihood probability distribution $P_1(f)$, see the orange line in Fig.~\ref{Fig3}~(b).
Then, according to Eq.~\eqref{Post}, we obtain the posterior $P_1^{Post}(f)$ and the posterior becomes the prior for the second iteration.
The voltage for the second iteration $U_2$ is determined by $P_1(f)$.
Since $P_{1}(f_d-\delta f)<P_{1}(f_d+\delta f)$, the central frequency is on the right side of $f_d$.
Thus, $s=0$ and $U_2$ should be increased according to Eq.~\eqref{Uk}.
$h$ controls the variation amplitude.
We choose $h_{min}=0.2$ and $h_{max}=2$ to restrict the minimum and maximum variation amplitude for simulation.
Similarly, the probability distributions for other iterations can be obtained step by step, see Fig.~\ref{Fig3}~(b)-(f).
Here, we choose $\delta U=2$ V, $\delta f=42$ kHz.

\subsection{Experimental demonstration}
Next, we perform the experimental demonstration.
Through implementing the adaptive Bayesian algorithm, we can obtain the suggested voltage for a desired magneto-sensitive transition frequency, as shown in Fig.~\ref{Fig4}.

Initially, there is no prior knowledge.
The mean and standard deviation for $N=1$ are calculated from a CPT spectrum of random initial voltage.
Generally, its standard deviation is large and its mean deviates severely from the target.
At this moment, we get a rough information for the target frequency and the applied voltage.
Continually, more CPT spectra for different applied voltage are observed in experiment.
Based on the earlier outcomes within the measurement sequence, the knowledge of the target frequency on the applied voltage can be gradually acquired.

\begin{figure}[!htp]
	\includegraphics[width=\columnwidth]{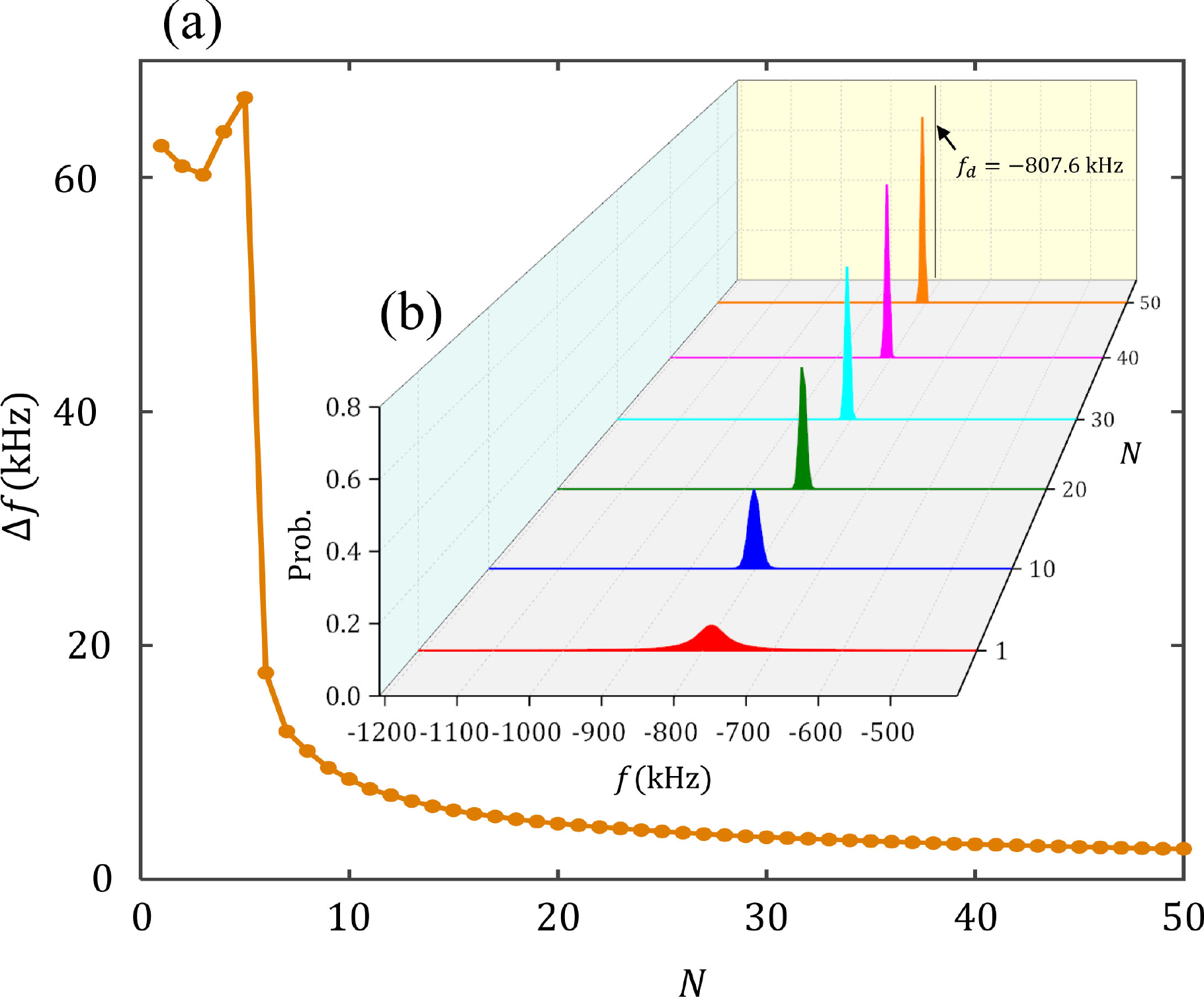}
	\caption{\label{Fig5} Convergence of standard deviation and posterior probability distribution. (a) The standard deviation $\Delta f$ versus the iteration number $N$. The converges for the mean and the standard deviation for the estimated frequency appear only after very limited iteration number. For a target frequency $f_d=-807.6$ kHz, starting from a random choice $U_0=8.5$ V, the standard deviation $\Delta f$ dramatically decreases after iteration number $N=5$. When $N=10$, $\Delta f$ becomes stable and gradually converges. (b) The posterior probability distribution for $N=1, 10, 20, 30, 40$ and $50$. Every posterior probability distribution contains the information from previous iterations. The multiple product among many similar probability distributions speeds up the convergence. As the iteration number increases, the posterior probability distribution becomes narrower and its mean value approaches the target frequency (denoted by the black solid line). }	
\end{figure}

As the iteration number becomes larger, more CPT spectra close to the target frequency are observed, the mean of the estimated frequency will locate closer and closer to the target frequency, and the voltage tends to be stable (see Fig.~\ref{Fig4}).
Meanwhile, the standard deviation of target frequency can be reduced dramatically after limited number of iteration [see Fig.~\ref{Fig5}~(a)].
Due to involving multiple products among many similar probability distributions, the posterior probability distribution becomes narrower when the iteration number increases, see Fig.~\ref{Fig5}~(b).
If the posterior probability after iteration is standardized to ranging from 0 to 1, a narrower CPT spectrum for the desired magneto-sensitive transition will be obtained.
The average of the voltage for the last few iteration number $U_d=\sum_{N=K}^{M}U_N/(M-K+1)$ can be used for achieving the desired magneto-sensitive transition at target frequency $f_d$.
Here, we choose the last ten iteration number, where $K=41$ and $M=50$.

Our method is simple and straightforward.
Before the experiment, one can know little about the relation between the applied voltage and the target frequency.
The initial voltage can be chosen randomly.
The random choice of initial conditions provides huge feasibility for practical applications, even in other experimental systems.
In addition, the standard deviation always dramatically decreases after limited iterations for different trials.
Here, our algorithm is verified in the standard case where the relation between the target frequency and the applied voltage is linear.

\begin{figure*}[!htp]
	\centering	
	\includegraphics[width=12cm]{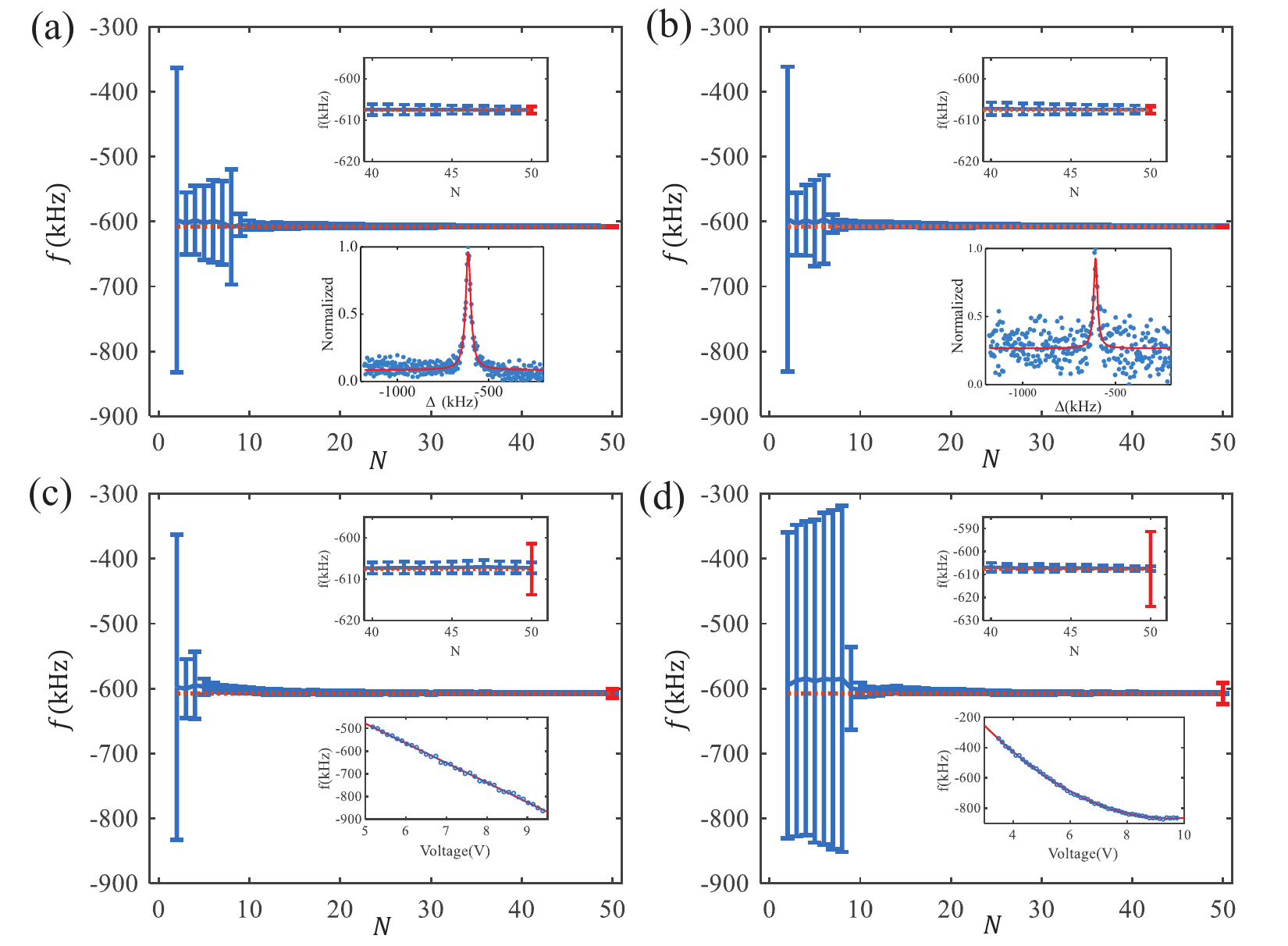}
	\caption{\label{Fig6} Experimental results via adaptive Bayesian algorithm under different scenarios. (a) Results obtained via normal magneto-sensitive CPT spectra. (b) Results obtained via noisy magneto-sensitive CPT spectra. (c) Results obtained via stochastic white noise of applied voltage. The linear relation between target frequency and applied voltage is shown in the inset. (d) Results obtained in the situation where the relation between the target frequency and the applied voltage is nonlinear. The red errorbars correspond to the standard deviations using traditional method extracted by fitting the relation $f_d=g(U_d)$ with the same number of measurement trials. The orange dashed lines denote the target frequency $f_d$=-607.6 kHz.}
\end{figure*}

To test the robustness of our algorithm, we take into account the noises in two different ways.
Firstly, we consider the observed CPT spectra become noisy, see the lower inset of Fig.~\ref{Fig6}~(b) whose signal-to-noise ratio of the CPT spectra is reduced compared to the one of Fig.~\ref{Fig6}~(a). Despite the CPT spectra become noisy, it can still converge to the target frequency and the corresponding standard deviation remains the same level after limited iterations compared to the case with normal CPT spectra as shown in Fig.~\ref{Fig6}~(a).
Secondly, we proactively add a small amount of stochastic white noise to each of the applied voltage, i.e., $U_k\rightarrow U_k+U_{\epsilon}$ where $U_{\epsilon}\in[-0.05 V,0.05 V]$ and $\overline{U_{\epsilon}}=0$. As shown in Fig.~\ref{Fig6}~(c), the applied voltage can still converge to the desired one. By contrast, using traditional method that manually extracting the relation $f_d=g(U_d)$ by fitting [see the inset of Fig.~\ref{Fig6}~(c)], the standard deviation (the red errorbar) is about 6.2 kHz, which is nearly 6 times larger than the one obtained via our algorithm. The experimental results show that our algorithm is more robust against different noises.

Furthermore, the advantage of our adaptive Bayesian algorithm becomes significant when the relation between the target frequency and the applied voltage becomes nonlinear.
To test its validity, we consider the situation where the target frequency $f_d$ is a quadratic function of the applied voltage $U_d$, i.e. $f \propto U^2$.
To get this relation, we set the voltage-controlled current as a quadratic function of $U_d$ via LabVIEW program.
As shown in inset of Fig.~\ref{Fig6}~(d), by manually varying the applied voltage and obtained the corresponding CPT spectra, one can roughly extract the relation $f_d=g(U_d)$ by fitting.
The obtained standard deviation is about 16.3 kHz (the red errorbar).
Instead, by using our algorithm, although the speed of convergence is not fast as the case of linear relation, the standard deviation can still dramatically decreases after iteration number $N=10$ [Fig.~\ref{Fig6}~(d)] and its standard deviation for $N=50$ reaches 1.1 kHz which is about 15 times smaller.
Our algorithm shows effectiveness and efficiency in dealing with the complicated cases where the relation between the target frequency and the controlled parameter is not linear.
\\

\section{Conclusions and discussions}

We have experimentally demonstrated how to automatically search a desired quantum transition by an adaptive Bayesian algorithm.
Our algorithm is based on the Bayes' theorem that the main features can be inferred effectively through updating the probability distribution after each measurement.
The adaptivity can effectively search the desired magneto-sensitive transition condition from a random initial one only after few iterations.
In particular, if the relation between the target frequency and the controlled parameter is nonlinear, our algorithm can be much more effective and efficient than the traditional methods.
Our adaptive Bayesian algorithm can be applied in many practical scenarios.

Firstly, this simple and efficient method for determining a desired transition frequency can be widely applied to precision frequency measurement such as developing practical CPT-based clocks~\cite{Ramsey1950,Ludlow2015,Sanner2018,Shuker2019}.
On one hand, the narrower posterior probability can improve spectral resolution.
On the other hand, our adaptive Bayesian algorithm can optimize the experimental conditions to reduce the systemic shifts, such as searching the zero magnetic field point.

Secondly, exchanging the roles of the transition frequency and the magnetic field, our CPT experiment with adaptive Bayesian algorithm can be inversely designed to probe an unknown static magnetic field $ B_s$.
For the magneto-sensitive transition we used, the target frequency $f_d=-2 \gamma B_z$ with $\gamma$ the gyromagnetic ratio of $^{87}$Rb atom~\cite{Schwindt2004}.
By implementing our adaptive Bayesian algorithm, the measured magnetic field $B_z=-f_d/(2\gamma)$ is the sum of the unknown static magnetic field $ B_s$ and the applied magnetic field $B_d$.
Since $B_d$ is proportional to $U_d$, the unknown static magnetic field can be deduced by the relation $ B_s=-f_d/(2\gamma)-B_d$.

Thirdly, the adaptive Bayesian algorithm can be widely extended to various physical systems where some certain controllable parameters can be introduced for adaptation.
Our algorithm may be improved to the cases of multiple controlled parameters.
The adaptation of controllable parameters can be modified according to the idea of gradient descent and so that the Bayesian method can further speed up the iteration procedure.
For more complicated cases, the experiment-design heuristics for the adaption can also be efficiently obtained via neural-network techniques~\cite{Puebla2020}.
\\
\\
\noindent\textbf{Acknowledgements}
\\
C. Han, J. Huang, and X. Jiang contribute to this work equally. This work is supported by the Key-Area Research and Development Program of Guangdong Province (2019B030330001), the National Natural Science Foundation of China (12025509, 11874434), and the Science and Technology Program of Guangzhou (201904020024). J.H. is partially supported by the Guangzhou Science and Technology Projects (202002030459). B.L. is partially supported by the Guangdong Natural Science Foundation (2018A030313988) and the Science and Technology Program of Guangzhou (201804010497).
\\
\\
\noindent\textbf{Author contributions}
\\
C.L., J.H. and B.L. conceived the project. C.H., J.H., X.J. and B.L. designed the experiment. J.H. and Y.Q. developed the theory. C.H., X.J., R. F. and B.L. performed the experiments. All authors discussed the results and co-wrote the manuscript, J.H. composed the first draft, C.L. wrote the final version. C.L. supervised the project.




\begin{thebibliography}{99}

\bibitem{Degen2017}
C. L. Degen, F. Reinhard, and P. Cappellaro,
\href{https://journals.aps.org/rmp/abstract/10.1103/RevModPhys.89.035002}{Rev. Mod. Phys. \textbf{89}, 035002 (2017).}

\bibitem{Mehta2019}
P. Mehta, M. Bukov, C. Wang, A. G. R. Day, C. Richardson, C. K. Fisher, and D. J. Schwab,
\href{https://doi.org/10.1016/j.physrep.2019.03.001}{Phys. Rep. \textbf{810}, 1124 (2019).}

\bibitem{Fiderer2021}
L. J. Fiderer, J. Schuff, and D. Braun,
\href{http://dx.doi.org/10.1103/PRXQuantum.2.020303}{PRX Quantum \textbf{2}, 020303 (2021).}

\bibitem{Li2018}
Y. Li, L. Pezz\`{e}, M. Gessner, W. Li, and A. Smerzi,
\href{https://www.mdpi.com/1099-4300/20/9/628}{ Entropy \textbf{20}, 628 (2018).}

\bibitem{Macieszczak2014}
K. Macieszczak, M. Fraas, and R. Demkowicz-Dobrzanski,
\href{https://iopscience.iop.org/article/10.1088/1367-2630/16/11/113002}{New J. Phys. \textbf{16}, 113002 (2014).}

\bibitem{Palittapongarnpim2019}
P. Palittapongarnpim and B. C. Sanders,
\href{https://journals.aps.org/pra/abstract/10.1103/PhysRevA.100.012106}{Phys. Rev. A 100, 012106 (2019).}

\bibitem{Becerra2013}
F. E. Becerra, J. Fan, G. Baumgartner, J. Goldhar, J. T. Kosloski and A. Migdall,
\href{https://www.nature.com/articles/nphoton.2012.316}{Nat. Photon. \textbf{7}, 147 (2013). }

\bibitem{Blume2010}
R. Blume-Kohout,
\href{https://iopscience.iop.org/article/10.1088/1367-2630/12/4/043034}{New J. Phys. \textbf{12}, 043034 (2010).}

\bibitem{Granade2012}
C. E. Granade, C. Ferrie, N. Wiebe, and D. G. Cory,
\href{https://iopscience.iop.org/article/10.1088/1367-2630/14/10/103013}{New J. Phys. \textbf{14}, 103013 (2012).}

\bibitem{Wang2017}
J. Wang, S. Paesani, R. Santagati, S. Knauer, A. A. Gentile, N. Wiebe, M. Petruzzella, J. L. O'Brien, J. G. Rarity, A. Laing, and M. G. Thompson,
\href{https://doi.org/10.1038/nphys4074}{ Nat. Phys. \textbf{13}, 551 (2017).}

\bibitem{Stenberg2014}
M. P. V. Stenberg, Y. R. Sanders, and F. K. Wilhelm,
\href{https://journals.aps.org/prl/abstract/10.1103/PhysRevLett.113.210404}{Phys. Rev. Lett. \textbf{113}, 210404 (2014).}

\bibitem{Granade2016}
C. Granade, J. Combes, and D. G. Cory,
\href{https://iopscience.iop.org/article/10.1088/1367-2630/18/3/033024}{New J. Phys. \textbf{18}, 033024 (2016).}

\bibitem{Ruster2017}
T. Ruster, H. Kaufmann, M. A. Luda, V. Kaushal, C. T. Schmiegelow, F. Schmidt-Kaler, and U. G. Poschinger,
\href{https://link.aps.org/doi/10.1103/PhysRevX.7.031050}{Phys. Rev. X \textbf{7}, 031050 (2017).}

\bibitem{Kaubruegger2021}
R. Kaubruegger, D. V. Vasilyev, M. Schulte, K. Hammerer, and P. Zoller,
\href{http://arxiv.org/abs/2102.05593}{arXiv:2102.05593.}

\bibitem{Wang2021}
G. Wang, D. E. Koh, P. D. Johnson, and Y. Cao,
\href{http://dx.doi.org/10.1103/PRXQuantum.2.010346}{PRX Quantum \textbf{2}, 010346 (2021).}

\bibitem{Dowling2003}
J. P. Dowling,  and G. J. Milburn,
\href{https://doi.org/10.1098/rsta.2003.1227}{Phil. Trans. R. Soc. Lond. A \textbf{361}, 1655 (2003).}

\bibitem{Puebla2020}
R. Puebla, Y. Ban, J. F. Haase, M. B. Plenio, M. Paternostro, and J. Casanova,
\href{https://arxiv.org/abs/2003.02151}{arXiv:2003.02151.}

\bibitem{Sarma2019}
S. D. Sarma, D. Deng, and L. Duan,
\href{https://doi.org/10.1063/PT.3.4164}{Physics Today \textbf{72}, 3, 48 (2019).}

\bibitem{Fosel2018}
T. F\"{o}sel, P. Tighineanu, T. Weiss, and F. Marquardt,
\href{https://link.aps.org/doi/10.1103/PhysRevX.8.031084}{Phys. Rev. X \textbf{8}, 031084 (2018).}

\bibitem{Friis2017}
N. Friis,  D. Orsucci, M. Skotiniotis, P. Sekatski, V. Dunjko, H. J. Briegel, and W. D\"{u}r,
\href{https://doi.org/10.1088/1367-2630/aa7144}{New. J. Phys. \textbf{19}, 063004 (2017).}

\bibitem{Rubio2019}
J. Rubio, and J. Dunningham,
\href{https://link.aps.org/doi/10.1103/PhysRevA.101.032114}{ Phys. Rev. A \textbf{101}, 032114 (2020).}

\bibitem{Lumino2018}
A. Lumino, E. Polino, Adil S. Rab, G. Milani, N. Spagnolo, N. Wiebe, and F. Sciarrino,
\href{https://link.aps.org/doi/10.1103/PhysRevApplied.10.044033}{ Phys. Rev. Applied \textbf{10}, 044033 (2018).}

\bibitem{Valeri2020}
M. Valeri, E. Polino, D. Poderini, I. Gianani, G. Corrielli, A. Crespi, R. Osellame, N. Spagnolo, and F. Sciarrino, \href{https://doi.org/10.1038/s41534-020-00326-6}{npj Quantum Information \textbf{16}, 92 (2020).}

\bibitem{Wiseman1995}
H. M. Wiseman,
\href{https://link.aps.org/doi/10.1103/PhysRevLett.75.4587}{ Phys. Rev. Lett. \textbf{75}, 4587 (1995).}

\bibitem{Berry2000}
D. Berry, and H. Wiseman,
\href{https://link.aps.org/doi/10.1103/PhysRevLett.85.5098}{ Phys. Rev. Lett. \textbf{85}, 5098 (2000).}

\bibitem{Higgins2007}
B. L. Higgins, D. W. Berry, S. D. Bartlett, H. M. Wiseman, and G. J. Pryde,
\href{https://doi.org/10.1038/nature06257}{ Nature \textbf{450}, 393 (2007).}

\bibitem{Armen2002}
M. A. Armen, J. K. Au, J. K. Stockton, A. C. Doherty,  and H. Mabuchi,
\href{https://link.aps.org/doi/10.1103/PhysRevLett.89.133602}{ Phys. Rev. Lett. \textbf{89}, 133602 (2002).}

\bibitem{Wheatley2010}
T. A. Wheatley, D. W. Berry, H. Yonezawa, D. Nakane,  H. Arao, D. T. Pope, T. C. Ralph, H. M. Wiseman, A. Furusawa,  and  E. H. Huntington,
\href{https://link.aps.org/doi/10.1103/PhysRevLett.104.093601}{ Phys. Rev. Lett. \textbf{104}, 093601 (2010).}

\bibitem{Hentschel2010}
A. Hentschel, and B. C. Sanders,
\href{https://link.aps.org/doi/10.1103/PhysRevLett.104.063603}{ Phys. Rev. Lett. \textbf{104}, 063603 (2010).}

\bibitem{Daryanoosh2018}
S. Daryanoosh, S. Slussarenko, D. W. Berry, H. M. Wiseman, and G. J. Pryde,
\href{https://doi.org/10.1038/s41467-018-06601-7}{ Nat. Commun. \textbf{9}, 4606 (2018).}

\bibitem{Bonato2017}
C. Bonato, and D. W. Berry,
\href{https://link.aps.org/doi/10.1103/PhysRevA.95.052348}{ Phys. Rev. A \textbf{95}, 052348 (2017).}

\bibitem{Nusran2014}
N. M. Nusran, and M. V. G. Dutt,
\href{https://link.aps.org/doi/10.1103/PhysRevB.90.024422}{ Phys. Rev. B \textbf{90}, 024422 (2014).}

\bibitem{Linden2014}
W. von der Linden, V. Dose, and U. von Toussaint,
\href{https://doi.org/10.1017/CBO9781139565608}{(Cambridge University Press,  2014).}

\bibitem{Jarzyna2015}
M. Jarzyna,  and R. Demkowicz-Dobrza\'{n}ski,
\href{https://doi.org/10.1088/1367-2630/17/1/013010}{ New J. Phys. \textbf{17}, 013010 (2015).}

\bibitem{Wiebe2016}
N. Wiebe,  and C. Granade,
\href{https://link.aps.org/doi/10.1103/PhysRevLett.117.010503}{ Phys. Rev. Lett. \textbf{117}, 010503 (2016).}

\bibitem{Dinani2019}
H. T. Dinani, D. W. Berry, R. Gonzalez, J. R. Maze,  and C. Bonato,
\href{https://link.aps.org/doi/10.1103/PhysRevB.99.125413}{ Phys. Rev. B \textbf{99}, 125413 (2019).}

\bibitem{Paesani2017}
S. Paesani,  A. A. Gentile, R. Santagati, J. Wang, N. Wiebe, D. P. Tew, J. L. O'Brien, and M. G. Thompson,
\href{https://link.aps.org/doi/10.1103/PhysRevLett.118.100503}{ Phys. Rev. Lett. \textbf{118}, 100503 (2017).}

\bibitem{Zibrov2010}
S. A. Zibrov, I. Novikova, D. F. Phillips, R. L. Walsworth, A. S. Zibrov, V. L. Velichansky, A. V. Taichenachev, and V. I. Yudin,
\href{https://link.aps.org/doi/10.1103/PhysRevA.81.013833}{ Phys. Rev. A \textbf{81}, 013833 (2010).}

\bibitem{Esnault2013}
F.-X. Esnault, E. Blanshan, E. N. Ivanov, R. E. Scholten, J. Kitching, and E. A. Donley,
\href{https://link.aps.org/doi/10.1103/PhysRevA.88.042120}{ Phys. Rev. A \textbf{88}, 042120 (2013).}

\bibitem{Liu2017}
X. Liu, E. Ivanov, V. I. Yudin , J. Kitching,  and E. A. Donley,
\href{https://link.aps.org/doi/10.1103/PhysRevApplied.8.054001}{ Phys. Rev. Applied \textbf{8}, 054001 (2017).}

\bibitem{Breschi2009}
E. Breschi, G. Kazakov, R. Lammegger, G. Mileti, B. Matisov, and L. Windholz,
\href{https://link.aps.org/doi/10.1103/PhysRevA.79.063837}{ Phys. Rev. A \textbf{79}, 063837 (2009).}

\bibitem{Mikhailov2010}
E. E. Mikhailov, T. Horrom, N. Belcher, and I. Novikova,
\href{http://josab.osa.org/abstract.cfm?URI=josab-27-3-417}{ J. Opt. Soc. Am. B \textbf{27}, 417 (2010).}

\bibitem{Hafiz2017}
M. A. Hafiz, G. Coget, P. Yun, S. Gu\'{e}erande, E. Clercq, and R. Boudot,
\href{https://doi.org/10.1063/1.4977955}{ J. Appl. Phys. \textbf{121}, 104903 (2017).}

\bibitem{Shahriar2014}
M. S. Shahriar, Y. Wang, S. Krishnamurthy, Y. Tu, G. S. Pati and S. Tseng,
\href{https://doi.org/10.1080/09500340.2013.865806}{ J. Mod. Opt. \textbf{61}, 351 (2014).}

\bibitem{Ramsey1950}
N. F. Ramsey,
\href{https://link.aps.org/doi/10.1103/PhysRev.78.695}{Phys. Rev. \textbf{78}, 695 (1950).}
	
\bibitem{Ludlow2015}
A. D. Ludlow, M. M. Boyd, J. Ye, E. Peik, and P. O. Schmidt,
\href{https://link.aps.org/doi/10.1103/RevModPhys.87.637}{Rev. Mod. Phys. \textbf{87}, 637 (2015).}

\bibitem{Sanner2018}
C. Sanner, N. Huntemann, R. Lange, C. Tamm, and E. Peik,
\href{https://link.aps.org/doi/10.1103/PhysRevLett.120.053602}{Phys. Rev. Lett. \textbf{120}, 053602 (2018).}

\bibitem{Shuker2019}
M. Shuker, J. W. Pollock, R. Boudot, V. I. Yudin, A. V. Taichenachev, J. Kitching, and E. A. Donley,
\href{https://link.aps.org/doi/10.1103/PhysRevLett.122.113601}{Phys. Rev. Lett. \textbf{122}, 113601 (2019).}

\bibitem{Schwindt2004}
P. D. D. Schwindt, S. Knappe, V. Shah, L. Hollberg, J. Kitching, L. A. Liew, and J. Moreland,
\href{https://doi.org/10.1063/1.1839274}{ Appl. Phys. Lett. \textbf{85}, 6409 (2004).}


\end{thebibliography}
\end{document}